\documentclass{article}
\usepackage{emulateapj,apjfonts}
\input psfig

\newcommand{\etal}{et~al.~}

\newcommand{\Msun}{M_\odot}
\newcommand{\mbh}{$M_\bullet$}

\newcommand{\kms}{$\rm {km}~\rm s^{-1}$}

\newcommand{\vdm}{van~der~Marel}
\newcommand{\degr}{$^\circ$}
\newcommand{\etnuk}{et~al.~}
\def\ts{\thinspace}

\def\mbh{$M_{\bullet}$~}

\newdimen\sa  \def\sd{\sa=.1em \ifmmode $\rlap{.}$''$\kern -\sa$
                               \else \rlap{.}$''$\kern -\sa\fi}

\begin{document}

\lefthead{Comparison of Black Hole Mass Estimates}

\righthead{Gebhardt~\etnuk}

\title{Black Hole Mass Estimates from Reverberation
Mapping and from Spatially Resolved Kinematics}

\author{Karl Gebhardt\altaffilmark{1,2,3}, John
Kormendy\altaffilmark{3}, Luis C. Ho\altaffilmark{4}, Ralf
Bender\altaffilmark{5}, Gary Bower\altaffilmark{6}, Alan
Dressler\altaffilmark{4}, S.M.~Faber\altaffilmark{2}, Alexei
V. Filippenko\altaffilmark{7}, Richard Green\altaffilmark{6}, Carl
Grillmair\altaffilmark{8}, Tod R. Lauer\altaffilmark{6}, John
Magorrian\altaffilmark{9}, Jason Pinkney\altaffilmark{10}, Douglas
Richstone\altaffilmark{10}, and Scott Tremaine\altaffilmark{11}}

\altaffiltext{1}{Hubble Fellow} 

\altaffiltext{2}{UCO/Lick Observatories, University of California,
Santa Cruz, CA 95064; gebhardt@ucolick.org, faber@ucolick.org}

\altaffiltext{3}{Department of Astronomy, University of Texas, Austin,
Texas 78712; gebhardt@astro.as.utexas.edu,
kormendy@astro.as.utexas.edu}
 
\altaffiltext{4}{The Observatories of the Carnegie Institution of
Washington, 813 Santa Barbara St., Pasadena, CA 91101;
lho@ociw.edu, dressler@ociw.edu}
 
\altaffiltext{5}{Universit\"ats-Sternwarte, Scheinerstrasse 1,
M\"unchen 81679, Germany; bender@usm.uni-muenchen.de}

\altaffiltext{6}{National Optical Astronomy Observatories, P. O. Box
26732, Tucson, AZ 85726; gbower@noao.edu, green@noao.edu,
lauer@noao.edu}

\altaffiltext{7}{Department of Astronomy, University of California,
Berkeley, CA 94720-3411; alex@astro.berkeley.edu}

\altaffiltext{8}{SIRTF Science Center, 770 South Wilson Ave.,
Pasadena, CA 91125; carl@ipac.caltech.edu}

\altaffiltext{9}{Institute of Astronomy, Madingley Road, Cambridge
CB3 0HA, England; magog@ast.cam.ac.uk}

\altaffiltext{10}{Dept. of Astronomy, Dennison Bldg., Univ. of
Michigan, Ann Arbor 48109; jpinkney@astro.lsa.umich.edu,
dor@astro.lsa.umich.edu}
 
\altaffiltext{11}{Princeton University Observatory, Peyton Hall,
Princeton, NJ 08544; tremaine@astro.princeton.edu}
 
\begin{abstract}

\pretolerance=10000  \tolerance=10000

Black hole (BH) masses that have been measured by reverberation
mapping in active galaxies fall significantly below the correlation
between bulge luminosity and BH mass determined from spatially
resolved kinematics of nearby normal galaxies.  This discrepancy has
created concern that one or both techniques suffer from systematic
errors. We show that BH masses from reverberation mapping are
consistent with the recently discovered relationship between BH mass
and galaxy velocity dispersion. Therefore the bulge luminosities are
the probable source of the disagreement, not problems with either
method of mass measurement. This result underscores the utility of the
BH mass -- velocity dispersion relationship. Reverberation mapping can
now be applied with increased confidence to galaxies whose active
nuclei are too bright or whose distances are too large for BH searches
based on spatially resolved kinematics.

\end{abstract}
 
\keywords{black hole physics -- galaxies: active -- galaxies: kinematics and
          dynamics -- galaxies: nuclei -- galaxies: Seyfert}

\section{Introduction}

\pretolerance=10000  \tolerance=10000

Searches for supermassive black holes (BHs) based on spatially
resolved kinematics have found $\sim$\ts35 candidates (see Kormendy
\etnuk 2000 for a review).  Almost all are in weakly active or
inactive galaxies.  The reason is that bright active galactic nuclei
(AGNs) swamp the light from the surrounding stars and gas, and
complicate the kinematic observations.  In addition, AGNs are rare, so
most are distant. Even with the {\it Hubble Space Telescope\/} ({\it
HST}), the central kinematics of galaxies are well enough resolved to
reveal BHs only in nearby galaxies.  The ironic result (Kormendy \&
Richstone 1995) is that the bright Seyfert nuclei and quasars that
motivate the BH search are conspicuously rare in the dynamical BH
census.

Reverberation mapping (Blandford \& McKee 1982; Netzer \& Peterson
1997) avoids this problem.  In this technique, time delays between
brightness variations in the continuum and in the broad emission lines
are interpreted as the light travel time between the BH and the
line-emitting region farther out. This provides an estimate of the
radius $r$ of the broad-line region (BLR).  We also have a velocity
$V$ from the FWHM of the emission lines.  Together, these measure a
mass $M_\bullet \approx V^2 r / G$, where $G$ is the gravitational
constant.  An important advantage is that the BLR is $\sim$\ts$10^2$
times closer to the BH than the stars and gas that are used in {\it
HST} spectroscopy.

However, several authors have pointed out that reverberation mapping
yields smaller BH masses at a given bulge luminosity than do dynamical
models of spatially resolved kinematics (e.g., Wandel 1999; Ho
1999). That is, in the observed correlation between BH mass and bulge
luminosity $L_{B,\rm bulge}$ (Kormendy 1993; Kormendy \& Richstone
1995), \mbh values from reverberation mapping are systematically low
(see Figure~1a which shows reverberation masses that are as much as a
factor of 40 smaller than predicted by the correlation). This
discrepancy is overstated when using the \mbh -- $L_{B,\rm bulge}$
correlation from Magorrian~\etnuk (1998).  Those BH masses are based
on two-integral models applied to low-resolution data. Comparison with
{\it HST} data and three-integral models shows that the Magorrian
\etnuk (1998) BH masses are high by about a factor of three, mainly
due to radially-biased anisotropy in the stellar orbits
(Gebhardt~\etnuk 2000c) that was not modeled in Magorrian~\etnuk
(1998) mass estimates.  Nevertheless, even using the best kinematic
data, Ho (1999) finds that \mbh values from reverberation mapping are
still low by a factor of $\sim5$ compared with masses based on
spatially resolved kinematics of different galaxies but similar bulge
luminosities.  It is important to resolve this discrepancy.


\begin{figure*}[t]
\centerline{\psfig{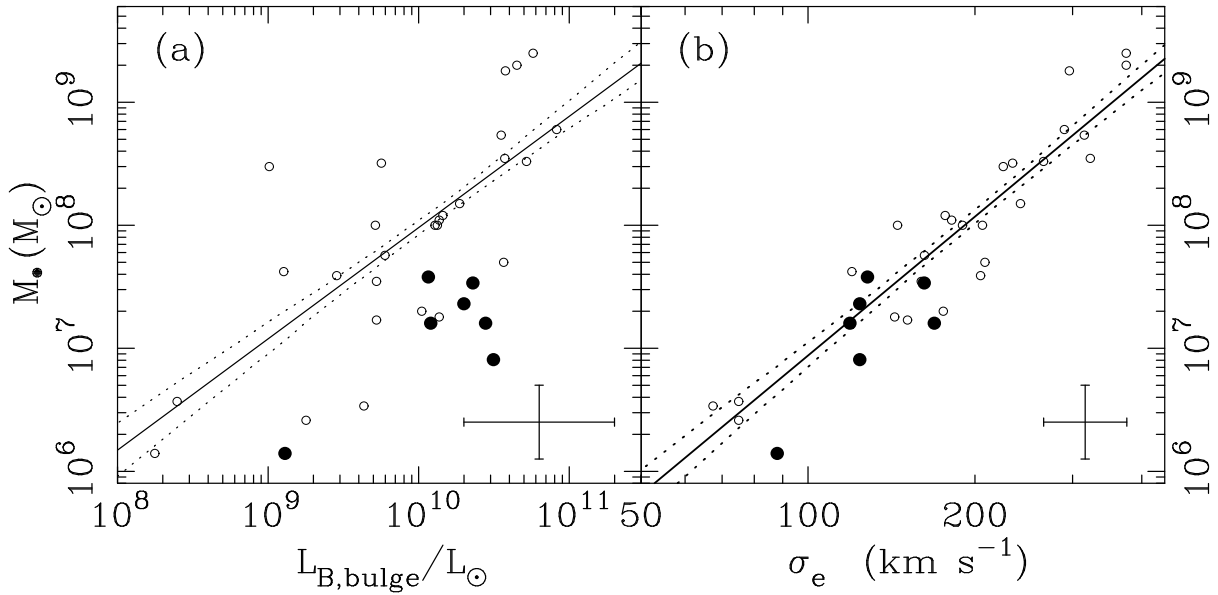}}
\figcaption[gebhardt.fig2.ps]{ Black hole mass versus (a) bulge
luminosity and (b) velocity dispersion. There are 33 points in the
dispersion plot: 26 from the compilation of Gebhardt~\etnuk (2000b)
(open circles), and seven points from reverberation mapping (filled
circles). Solid and dotted lines are the best-fit correlations and
their 68\% confidence bands from Gebhardt~\etnuk (2000b) fitted only
to the galaxies with spatially resolved kinematics. The error bars in
the lower right for each plot are representative for the reverberation
mapping uncertainties.
\label{fig1}}
\end{figure*}


Gebhardt~\etnuk (2000b) and Ferrarese~\& Merritt (2000) find a new
correlation between \mbh and the effective velocity dispersion
$\sigma_e$ of the host galaxy. This relation is significantly tighter
than the \mbh -- $L_{B,\rm bulge}$ correlation, consistent with zero
intrinsic scatter. In this {\it Letter}, we add reverberation mapping
masses to the new correlation and find that the systematic offset
between the two mass estimators is no longer significant.

\section{Reverberation Masses and Velocity Dispersions}

Ho (1999) and Wandel, Peterson, \& Malkan (1999) measure \mbh
values for 22 Seyfert~1 galaxies using reverberation
mapping. Unfortunately, the absorption-line kinematics of these
galaxies are not well studied, so we are unable to obtain velocity
dispersions for the whole sample. Only seven galaxies have usable
published dispersions. The three sources for these dispersions are
Nelson~\& Whittle (1995), Di~Nella~\etal (1995), and Smith, Heckman,
\& Illingworth (1990).

For most of these galaxies, the velocity dispersions are difficult to
measure. Some are late-type galaxies, so template matching is
difficult because of the presence of young stars.  In many cases,
dilution of the stellar absorption lines by the nonstellar continuum
of the AGN is a problem. Dilution does not alter the velocity
dispersion of the lines, but it does make them hard to detect.
Ideally, we should use spectral regions that are minimally sensitive
to template mismatch and to line dilution.  The calcium infrared
triplet near 8500 \AA\ is preferable to the traditional Mg~b
$\lambda$5170 region (Dressler 1984).  In the present paper, we adopt
velocity dispersions derived from the calcium triplet region whenever
possible.  

The study of Terlevich, D\'\i az, \& Terlevich (1990) contains three
galaxies with reverberation masses; however, the dispersions measured
for many of their other galaxies do not compare well with those from
other groups. For example, their dispersions for M33, M32, and M31 are
significantly different than the accepted values in their apertures:
77~\kms\ compared with 21~\kms\ (Kormendy~\& McClure 1993) for M33,
56~\kms\ compared with $\sim$ 80~\kms\ (van der Marel~\etal 1994) for
M32, and 137~\kms\ compared with 195~\kms\ (van der Marel~\etal
1994) for M31. Therefore we exclude their measurements from our
analysis.

For their dispersion estimate, Gebhardt~\etnuk (2000b) use the
projected, luminosity-weighted value inside the half-light or
effective radius of the bulge, which we call the effective dispersion
and denote by $\sigma_e$. For the AGN sample, we do not have
dispersion profiles and cannot perform the same
calculation. Consequently, we must use {\it central}
dispersions. However, based on their sizes, these galaxies have bulge
half-light radii of only a few arcseconds (Kotilainen, Ward, \&
Williger 1993; Baggett, Baggett, \& Anderson 1998). These sizes are
similar to the typical seeing and extraction window used
($\sim$2\arcsec).  Gebhardt~\etnuk (2000b) find that central aperture
dispersions measured at this resolution are similar on average to
effective dispersions, with a scatter of at most 10\%. Thus, the
reported dispersion should be a good approximation to the effective
dispersion, although a systematic study using the dispersion profile
would be worthwhile. When using central dispersions, the most crucial
concern is whether the black hole affects the measured
dispersion. Assuming typical stellar mass-to-light ratios, the spheres
of influence for these black holes are a few tenths of an arcsecond.
They should have little effect on the dispersions.

The effective dispersion used in Gebhardt~\etnuk (2000b) assumes
edge-on configuration, and since the projected dispersion varies with
orientation we must consider whether it needs correction for
inclination. This effect may be more important in AGN disk galaxies,
where we might expect significant rotation (a rotating galaxy will
have a larger projected dispersion edge-on than face-on). Four of the
seven galaxies are inclined greater than 45\degr\ and are likely to
have corrections smaller than their uncertainties. The three galaxies
more face-on than 45\degr\ are Mrk\,590, NGC\,4151, and
NGC\,4593. Based on systems with large bulge fractions, even these
would have corrections less than 10\% (Gebhardt~\etnuk 2000b);
however, better kinematic data on the bulge rotation profiles for a
larger sample of AGNs is needed before we fully understand inclination
corrections.

Ho~(1999) compiled bulge luminosities and bulge-to-total light ratios
($B/T$) from three sources. Kotilainen~\etal (1993) provide surface
photometry and disk--bulge decompositions for 3C\,120, Mrk\,590,
NGC\,3227, NGC\,4151, and NGC\,4593; Granato~\etal (1993) provide
disk--bulge decompositions for Mrk\,590 and NGC\,3516; while
Baggett~\etal (1998) give profiles and decompositions for NGC\,3227,
NGC\,4051, NGC\,4151, and NGC\,4593. For the galaxies that overlap
among the various groups, we find consistent $B/T$ values. However,
each study uses a de~Vaucouleurs profile for the bulge component.  If
these bulges are more nearly exponential or if the AGN contributes
significant light unaccounted for, then the bulge light will have been
overestimated.

Table~1 lists the data we have discussed, and Figure~1 plots \mbh
versus the bulge luminosity and effective dispersion $\sigma_e$. The
masses from resolved kinematics and the associated least-squares fits
come from Gebhardt~\etnuk (2000b). The relation fitted only to the
galaxies with spatially resolved kinematics for the \mbh -- $\sigma_e$
correlation is \mbh $ = 1.2 (\pm0.2)\times10^8 M_\odot
(\sigma_e/200\hbox{\,\kms})^{3.75\,(\pm0.3)}$.  The reverberation
masses lie a factor of 5--10 too low in the luminosity plot (black
dots), but are much more consistent with the correlation in the
$\sigma_e$ plot. In the $\sigma_e$ relation, the reverberation mapping
masses have an average offset of $-0.21$ ($\pm0.13$) dex and a
dispersion of 0.34 dex relative to that average. The scatter (0.30 dex
in log\,$M_\bullet$ at fixed dispersion) is the same regardless of
whether or not we include the reverberation mapping masses in the fit,
but the slope changes from 3.75 to 3.90 if we include them.

The average uncertainties in M$_\bullet$, $L_B$, and $\sigma_e$ for
the reverberation mapping estimates are shown in the bottom corner in
Fig.~1. Unfortunately, the uncertainties in reverberation mapping are
dominated by systematics that are uncertain or unknown (Wandel~\etal
1999) and can be quite large. Since we have only seven reverberation
mapping masses, we do not attempt a rigorous statistical analysis
including the measurement uncertainties.

\newcommand{\spa}{\phantom{1}}
\newcommand{\spb}{\phantom{11}}
\begin{deluxetable}{llcccccc}
\tablenum{1}
\tablewidth{36pc}
\tablecaption{Seyfert Galaxies with Reverberation Mapping BH Masses}
\tablehead{
\colhead{Galaxy}              & 
\colhead{Type}                &
\colhead{$D$}                 &
\colhead{$L_{B,{\rm bulge}}$} &
\colhead{$B/T$}               &
\colhead{$M_\bullet$}         &
\colhead{$\sigma$}            &
\colhead{Source for $\sigma$} \nl
\colhead{}                    & 
\colhead{}                    &
\colhead{(Mpc)}               &
\colhead{($10^{10}\,L_{\odot}$)}&
\colhead{}                    &
\colhead{($\Msun$)}           &
\colhead{(\kms)}              &
\colhead{}                    }
\startdata
 3C\,120   & S0:   & 132  & 2.29 & 0.24 & $3.4\times10^7$ &  162  & Smith~\etal 1990      \nl
 Mrk\,590  & Sa:   & 105  & 2.78 & 0.47 & $1.6\times10^7$ &  169  & Nelson~\& Whittle 1995\nl
 NGC\,3227 & SABa  &\spa21& 1.16 & 0.52 & $3.8\times10^7$ &  128  & Nelson~\& Whittle 1995\nl
 NGC\,3516 & SB0:  &\spa39& 2.00 & 0.61 & $2.3\times10^7$ &  124  & Di~Nella~\etal    1995\nl
 NGC\,4051 & SABbc &\spb9 & 0.13 & 0.20 & $1.4\times10^6$ &\spa88 & Nelson~\& Whittle 1995\nl
 NGC\,4151 & SABab &\spa20& 1.20 & 0.36 & $1.6\times10^7$ &  119  & Nelson~\& Whittle 1995\nl
 NGC\,4593 & SBb   &\spa40& 3.13 & 0.48 & $8.1\times10^6$ &  124  & Nelson~\& Whittle 1995\nl
\enddata
\end{deluxetable}

\section{Discussion}

The apparent discrepancy between reverberation mapping and dynamical
modeling of spatially resolved kinematics arose from a comparison of
\mbh with bulge luminosities. Since the reverberation mapping masses
are consistent with the \hbox{\mbh -- $\sigma_e$} correlation and not
with the \mbh -- $L_{B,\rm bulge}$ correlation, the discrepancy in the
latter is likely due to problems with the use of bulge luminosities,
not with estimation of the BH masses.  Velocity dispersions are more
difficult to measure in AGNs, but we have little reason to suspect
that they have systematic errors.

However, it is important to examine the potential complications of
both techniques.  Sections 3.1 and 3.2 below suggest that the BH
masses from resolved stellar kinematics have only small systematic
errors but that the BH masses from reverberation mapping may be biased
slightly low.

\subsection{Complications in the Stellar Dynamical Samples}

      (1) Model limitations were once a concern but are now under
control.  The current state of the art is to use Schwarzschild's
method (Schwarzschild 1979; Richstone \& Tremaine 1988) to construct
three-integral models that include galaxy flattening and velocity
anisotropy (van der Marel~\etal 1998; Gebhardt \etnuk 2000a;
Richstone~\etnuk 2000). The galaxies with stellar kinematical masses
in Figure 1a,b all have three-integral models. When such models are
fitted to {\it HST} data, the errors in $M_\bullet$ are
small. However, there is still some concern about whether
non-axisymmetric structure affects the masses, and thorough
comparisons of the different modeling codes have not yet been carried
out.

    (2) Selection effects may be present since early BH searches were
biased toward objects with unusually high BH masses (the first {\it
HST} targets were galaxies that showed high central dispersions at
ground-based resolution). This bias still persists in the current
overall BH census based on stellar-dynamical measurements, but the
present sample is large enough to overcome effects from a few galaxies
with high BH masses.

    (3) It is possible that galaxies contain central concentrations of
ordinary dark matter (e.{\ts}g., stellar remnants) that are included
in most BH mass measurements.  This concern is prompted by the fact
that the radii that we resolve with {\it HST} spectroscopy are
$\sim10^2$ larger than the BLR that is used in reverberation
mapping. However, measurements in our Galaxy (Genzel~\etal 1997, 2000;
Ghez~\etal 1998) and in NGC\,4258 (Greenhill~\etal 1996) probe a small
region comparable to that probed by reverberation mapping in
other galaxies, and find no suggestion of any dark mass in addition to
a BH.

\subsection{Complications in the Reverberation Mapping Samples}

    (1) The geometry and orbital distribution of the BLR are poorly
known. If, as is often assumed in the AGN unification model (Antonucci
1993), Seyfert~1 nuclei are viewed preferentially face on and if the
BLR and the obscuring tori are roughly coplanar, then the inclination
correction for reverberation masses would be significant. However,
because the thickness of the BLR disk is unknown, the actual
correction is uncertain. Nonetheless, if the corrections are factors
around 2, then the comparison of these masses in Fig.~1b will improve.

    (2) The measured ``lag'' in reverberation studies is a peculiar
moment over the distribution of distances between the central nucleus
and the line-emitting gas, and the measured line width is a different
peculiar moment over the velocity distribution. These are affected by
the adopted weightings, the shape of the continuum fluctuation power
spectrum, and the sampling of the monitoring.

    (3) Selection effects restrict the BH masses that are currently
measurable by reverberation mapping.  The timescales of AGN
variability scale with luminosity (e.g., Netzer \& Peterson 1997) and
presumably with mass for Eddington-limited systems.  Ongoing studies
of slowly-varying, high-luminosity AGNs (Kaspi \etal 2000) and of
rapidly-varying, low-luminosity AGNs (Peterson \etal 2000) should
remedy this situation in the future. 

    (4) It is important to consider non-gravitational effects acting
on BLR gas.  They include radial motions caused by radiation pressure
or by mechanical energy from jets.  The resulting mass measurement
errors could have either sign, but it is most likely that we would
overestimate $M_\bullet$ (Krolik 1997).

The following are additional complications that arise when estimating
bulge luminosities.

    (5) AGNs are commonly associated with starbursts (e.{\ts}g.,
Heckman 1999; Sanders 1999).  This may cause $M_\bullet$ to look too
small in the $M_\bullet$ -- $L_{B,\rm bulge}$ correlation. Although
$M_\bullet$ is plotted against blue luminosity, the physical
correlation is presumably with bulge mass, and star formation can
easily reduce $M/L_{B}$ by a factor of 2 -- 4 compared to its value in
bulges that are made of old stars.  Then the bulge would look too
bright for the given $M_\bullet$.

    (6) It is possible that the light from the AGN biases the estimate
of the bulge light. First, the AGN makes the center of the galaxy look
exceptionally bright in poor-quality images, so there is a tendency to
assign a Hubble type that is too early if using qualitative visual
inspection. If one then uses the loose correlation between Hubble type
and bulge-to-disk ratio (Simien \& de Vaucouleurs 1986) to estimate
bulge luminosities, they will be overestimated. Second, even if one
uses disk/bulge decompositions, unless the AGN is modeled separately,
it will likely cause an overestimate of the bulge light as well.

\section{Conclusion}

We have shown that masses derived from reverberation mapping are
consistent with the relation between BH mass and galaxy velocity
dispersion derived from spatially resolved kinematics. Based on a
sample of seven Seyfert galaxies, we find that the systematic and
random errors in BH masses determined from reverberation mapping are
around 0.21 dex and 0.34 dex, respectively. It is remarkable that,
despite the large number of possible systematic biases (especially in
reverberation mapping), both methods appear to provide consistent and
reliable estimators of BH masses. Peterson \& Wandel (2000) provide
further support of the reliability for the AGN mass estimates by
showing a Keplerian relation between line width and time lag. One
could even use the $M_\bullet-\sigma_e$ correlation to infer
properties of the broad-line region; for example, any differences in
the AGN masses compared with the correlation may provide insight into
the BLR geometry.

The fact that reverberation mapping successfully delivers BH masses
offers tremendous hope of getting BH masses in objects that otherwise
would not be accessible, namely bright AGNs, including QSOs (e.g.,
Kaspi~\etal 2000), and high-redshift AGNs. The latter hold some hope
of probing the time evolution (growth history) of BH mass (e.g.,
Wandel 1999).  Furthermore, the correlation between photoionization
and reverberation models (Wandel~\etal 1999) offers the possibility of
wholesale AGN mass estimates. Future studies aimed at comparing the
two mass estimators on the {\it same} galaxies are required to confirm
both techniques, but the present results are encouraging.

\acknowledgements

We are grateful for comments from B. Peterson, A. Wandel, J. Krolik,
and the referee, K. Anderson. This work was supported by {\it HST}
grants to the Nukers, GO--02600.01--87A, G06099, and G07388, and by
NASA grant NAG5-8238. A.V.F. acknowledges NASA grant
NAG5-3556. K.G. is supported by NASA through Hubble Fellowship grant
HF-01090.01-97A awarded by the Space Telescope Science Institute,
which is operated by the Association of the Universities for Research
in Astronomy, Inc., for NASA under contract NAS 5-26555.

\end{document}